\documentclass[10pt,twocolumn]{article}
\usepackage[margin=1.8cm]{geometry}
\usepackage[T1]{fontenc}
\usepackage{times}
\usepackage{titlesec}

\usepackage[round]{natbib}
\setcitestyle{aysep={}}
\usepackage{graphicx,float,caption,subcaption,amsmath}
\usepackage{amssymb}
\usepackage{dblfloatfix} 

\graphicspath{{./}{/mnt/data/}}
\DeclareGraphicsExtensions{.pdf,.png,.jpg,.jpeg}
\newcommand{\img}[2][]{\includegraphics[#1]{\detokenize{#2}}}

\usepackage{indentfirst}
\titlespacing*{\section}{0pt}{0.8ex plus .2ex minus .2ex}{0.8ex plus .2ex minus .2ex}

\title{\bfseries Push-response anomalies in high-frequency S\&P 500 price series}
\author{%
  Dmitrii Vlasiuk \quad Mikhail Smirnov\\
  {\small Department of Mathematics, Columbia University}\\
  {\small November 2025}
}
\date{}

\begin{document}

\twocolumn[
\maketitle

\begin{center}
\begin{minipage}{\textwidth}
\begin{abstract}
We test the hypothesis that consecutive intraday price changes in the most liquid U.S. equity ETF (SPY) are conditionally nonrandom. Using NBBO event-time data for about 1{,}500 regular trading days, we form for every lag \(L\) ordered pairs of a backward price increment (``push'') and a forward price increment (``response''), standardize them, and estimate the expected responses on a fine grid of push magnitudes. The resulting lag-by-magnitude maps reveal a persistent structural shift: for short lags (1-5{,}000 ticks), expected responses cluster near zero across most push magnitudes, suggesting high short-term efficiency; beyond that range pronounced tails emerge, indicating that larger historical pushes increasingly correlate with nonzero conditional responses, in defiance of random behavior. We also find that large negative pushes are followed by stronger positive responses than equally large positive pushes, which is consistent with asymmetric liquidity replenishment after sell-side shocks. Decomposition into symmetric and antisymmetric components and the associated dominance curves confirm that short-horizon efficiency is restored only partially. The evidence points to an intraday, lag-resolved anomaly that is invisible in unconditional returns and that can be used to define tradable pockets and risk controls.
\end{abstract}
\end{minipage}
\end{center}
\vspace{0.8em}
] 

\section{Introduction}

The Efficient Market Hypothesis developed by Samuelson (1965) and Fama (1965, 1970) suggests that financial markets are efficient. Nevertheless, the presence of anomalies -- persistent and predictable patterns in asset price behavior that suggest the empirical regularity of returns beyond random fluctuation -- provides a foundation for active trading strategies which, if robust to model misspecification and implementation costs, can deliver stable performance.

SPY is an exchange-traded fund tracking the S\&P 500 index. Previous empirical studies, such as Martin (2021), have provided evidence that in the long term SPY exhibits efficient behavior, yet more advanced data analysis techniques borrowed from statistical physics and related quantitative disciplines reveal patterns that remain invisible to standard tests. Moreover, due to the proprietary nature of the quantitative finance field, it is often the case that a large portion of research remains undisclosed. Therefore, this paper is motivated by the opportunity to push the boundaries of standard statistical analysis of financial price series and focused on the exploration of the autocorrelative anomalies exhibited by the SPY top‑of‑the‑book data.

While the random walk view emphasizes informational efficiency at daily horizons, the microstructure literature shows that intraday returns can display systematic patterns induced by trading frictions. Roll (1984), Glosten and Harris (1988), and Hasbrouck (1991) demonstrate how bid-ask bounce, inventory control, and adverse selection generate short-horizon negative serial correlation and strong covariation between price changes and order imbalances. Biais, Hillion, and Spatt (1995) link these effects to the way liquidity is supplied and demanded in the limit order book, and Lee and Ready (1991) make the relevant order-flow signal observable through trade-sign classification. These results motivate our design: condition on a realized push in order flow or price and test for a predictably signed, state-dependent response over the next few thousand events.

At the market level, Chordia, Roll, and Subrahmanyam (2005) show that order-imbalance persistence coexists with near-zero daily return autocorrelation, implying rapid intraday adjustment back toward efficiency. As a unifying reference, Bouchaud, Bonart, Donier, and Gould (2018) provide evidence of long memory in order flow and of transient, state-dependent price impact. In this light, our lag-resolved push-response surface yields a nonparametric estimate of the impact propagator as a function of standardized push magnitude and horizon.

In parallel to these microstructure findings, Kyle (1985) showed that when informed and liquidity-driven trades arrive jointly, prices adjust in a signed and state-dependent way, such that a single trade innovation can have a transient and a permanent component. Madhavan, Richardson, and Roomans (1997) documented that quote revisions can be explained by an explicit mixture of public-information shocks, inventory rebalancing, and order-processing costs, which implies that consecutive price increments need not be fully random even in highly liquid markets. Huang and Stoll (1997) provided a general decomposition of the bid-ask spread into adverse-selection and inventory parts and showed that these parts can dominate at different horizons. 

Intraday predictability also arises in a systematic pattern: Heston, Korajczyk, and Sadka (2010) show that returns tend to repeat at half-hour intervals aligned with the trading day, which is consistent with execution flow and institutional trading schedules. Optimal-execution and market-impact models provide a complementary structural view of this mechanism. Almgren and Chriss (2001) and Almgren, Thum, Hauptmann, and Li (2005) model how specified execution schedules and impact functions translate order flow into prices, while Gatheral (2010) shows that no-dynamic-arbitrage restrictions sharply limit the admissible shape and decay of transient price impact. Taken together, these results motivate the push-response design we study, in which measurable shocks in order flow are linked to subsequent intraday price adjustments.

Our construction follows the two-increment conditioning idea studied by Leonidov, Trainin, Zaitsev, and Zaitsev (2006a, 2006b, 2006c, 2007, 2009), who documented asymmetric “market mill” geometry between a backward push and a forward response. We extend that idea in three directions: (i) we embed it in a lag-resolved 3D surface (lag × standardized push) rather than a single-lag or few-scale view; (ii) we estimate a full heatmap and decompose it into symmetric and antisymmetric parts with an explicit dominance measure; and (iii) we scale estimation to NBBO event time with far more lags and standardized grids so that cross-lag structure is directly comparable.

In the paper, rather than observing and analyzing aggregate autocorrelation, we propose the push-response decomposition, where pushes represent the initial price change at a fixed lag and responses are the corresponding price change that follows. Then, we test whether responses are systematically directional or magnitude-driven at given horizons. The working hypotheses are as follows: at some lags the sign of the push aligns with the sign of the response, which indicates momentum. At other lags, the signs oppose, which indicates reversal. The balance between these regimes varies with $L$, which represents the lag, and with $|z_p|$, which represents the normalized push. 

The outline of the paper is as follows. Section II discusses the applied data and preprocessing techniques. Then, in Section III, the push-response decomposition method is outlined and further analysis and diagnostics tools, including local dominance measurement and symmetry and antisymmetry analysis, are discussed. Finally, Section IV includes the 3D push-response-lag surfaces, symmetry heatmap, and local and aggregate dominance curves. Section V contains a further discussion and implications for real-life trading. Section VI contains conclusions and potential for further research.

\section{Data and Preprocessing}

The analysis is based on the National Best Bid and Offer (NBBO) quote data provided by TAQ for the SPDR S\&P~500 ETF Trust (SPY) covering the period from January~2018 to December~2023.  
For each event, the mid price was computed as the arithmetic mean of the best bid and best ask prices.  
The mid price is preferred to the transaction price because it reflects the instantaneous consensus of supply and demand without being affected by short-term trade direction or bid-ask bounce.  
To ensure that only firm quotes contributed to the series, the dataset was restricted to records carrying the \textit{R} condition flag, which indicates regular quotes eligible for NBBO formation under Regulation~National Market System.

\vspace{0.4em}
We construct the NBBO from quotes across the seven major US equity exchanges: NYSE, NASDAQ, NYSE~Arca, Cboe~BZX, Cboe~BYX, Cboe~EDGX, and Cboe~EDGA.  
These venues were chosen because they consistently constitute the dominant share of displayed liquidity for SPY and provide continuous, high-quality quote updates.  
The inclusion of multiple exchanges allows the reconstruction of the consolidated best bid and offer with minimal fragmentation noise.  
Top-of-the-book quotes were extracted for each of these venues and merged to produce the NBBO sequence for every event.

\vspace{0.4em}
Data quality control was carried out in several steps.  
First, the quote stream was ordered in event time, meaning that each observation represents one update to the NBBO rather than a fixed clock interval.  
Event time avoids the unequal sampling problem inherent in clock time and ensures that each increment corresponds to an actual information event.  
Only records from the Regular Trading Hours (09:30–16:00 ET) session were retained in order to exclude overnight and pre-market periods, whose low liquidity and sporadic updates distort the empirical distribution of price changes.

\vspace{0.4em}
Extreme values were mitigated by applying a winsorization procedure to the sequence of mid-price returns.  
Let $r_t$ denote the mid-price increment at event $t$.  
Define the winsorized series $\tilde{r}_t$ as

\begin{equation}
\tilde{r}_t = 
\begin{cases}
q_{0.00001}, & r_t < q_{0.00001}, \\
r_t, & q_{0.00001} \le r_t \le q_{0.99999}, \\
q_{0.99999}, & r_t > q_{0.99999}, \tag{2.1}
\end{cases}
\end{equation}

where $q_{p}$ denotes the empirical $p$-quantile of $\{r_t\}$.  
This procedure trims the most extreme 0.001\% of observations in each tail of the return distribution, reducing the influence of outliers that arise from potential erroneous quotes while preserving genuine market movements.

\vspace{0.4em}
After winsorization, the residual jumps larger than \$1.50 inside a single trading day were removed.  
Such discontinuities are inconsistent with the continuous double-auction mechanism in a highly liquid instrument such as SPY and almost invariably correspond to erroneous aggregated ticks or exchange outages.  
Formally, any pair of consecutive mid prices $m_t$ and $m_{t-1}$ satisfying

\begin{equation}
|m_t - m_{t-1}| > 1.5 \tag{2.2}
\end{equation}

was treated as a data error, and both affected events were excluded from the sample, which still preserved over 99.5\% of the data.

\vspace{0.4em}
The resulting dataset comprises approximately $2.6\times10^{9}$ valid NBBO events spanning about 1,512 regular trading days.  
The combination of event-time sampling, exchange-level NBBO reconstruction, and outlier control yields a well-constructed representation of SPY price formation suitable for subsequent lag-resolved push-response analysis without significant distortions of the price series.

Our cleaning procedure is consistent with the high-frequency volatility literature. Andersen, Bollerslev, Diebold, and Labys (2003) showed that realized-volatility measures built from tick-level data are highly sensitive to even a small fraction of contaminated observations; Barndorff-Nielsen and Shephard (2002) reached the same conclusion in the context of nonparametric volatility estimation. Ait-Sahalia, Mykland, and Zhang (2011) and Ait-Sahalia and Xiu (2019) further demonstrated that when microstructure noise is present, one must either filter or explicitly model the noise to recover the latent continuous-time price. Because our goal is to estimate \emph{conditional} expectations of forward returns at many lags, even rare abnormal ticks would bias the tails of the binned push distribution, so we follow these authors in trimming extremes and enforcing regular-trading-hours continuity.

\section{Methodology}

Gatheral (2010) derived no-dynamic-arbitrage restrictions on admissible impact functions, and in our context this means that any nonzero conditional response we detect should be localized in lag-magnitude space and should level off instead of increasing without bound. It is important to note that we do not impose no-arbitrage conditions initially and aim to look at the output produced by the data. Localized deviations from no-arbitrage conditions would imply local anomalies, and in that case they will be discussed separately.

\subsection{Definitions and pair construction}

Engle and Russell (1998) argued that for high-frequency data it is often more appropriate to index observations in event time rather than clock time, because information arrives with quote updates. For this reason we form triplets \((t-L, t, t+L)\) out of event time only when all three belong to the same regular session. This keeps the push and the response on the same event-time scale.

For a given nonnegative integer lag $L$, we define the \emph{push} and \emph{response} at anchor $t$ representing event time as raw price changes
\begin{equation}
p_t(L) = m_t - m_{t-L}, \tag{3.1}
\end{equation}
\begin{equation}
r_t(L) = m_{t+L} - m_t. \tag{3.2}
\end{equation}

We form the ordered pair $(p_t(L), r_t(L))$ only when the three anchors $t-L$, $t$, and $t+L$ all lie within the same regular trading session. Pairs that would cross the open or the close are not constructed.

The construction in Bouchaud, Gefen, Potters, and Wyart (2004) treats a signed order-flow shock as producing a price impact that is strong at short horizons and decays as the horizon increases. In our setting the push \(p_t(L)\) plays the role of this signed shock and the response \(r_t(L)\) is the subsequent impact, but we estimate the relation nonparametrically for many lags \(L\) on a common grid instead of fitting a single decay function.

We study two families of lags. The short family targets microsecond-to-minute structure:
\begin{equation}
L \in \{1\} \cup \{50, 100, 150, \dots, 5000\}. \tag{3.3}
\end{equation}
The long family targets second-to-multi-minute structure:
\begin{equation}
L \in \{1000, 2000, 3000, \dots, 500000\}. \tag{3.4}
\end{equation}
For every $L$ in either family we enforce the same-session constraint stated above, so both $p_t(L)$ and $r_t(L)$ are computed strictly within a single RTH day.

Volatility scales with $L$, so raw values are not comparable across horizons. For each $L$ we estimate the lag-specific first and second moments using all admissible anchors,
\begin{equation}
\mu_p(L) = \mathbb{E}[\,p_t(L)\,], \qquad \sigma_p(L) = \sqrt{\mathbb{V}\mathrm{ar}[\,p_t(L)\,]}, \tag{3.5}
\end{equation}
\begin{equation}
\mu_r(L) = \mathbb{E}[\,r_t(L)\,], \qquad \sigma_r(L) = \sqrt{\mathbb{V}\mathrm{ar}[\,r_t(L)\,]}. \tag{3.6}
\end{equation}
We then form standardized variables
\begin{equation}
z_p = \frac{p_t(L) - \mu_p(L)}{\sigma_p(L)}, \qquad z_r = \frac{r_t(L) - \mu_r(L)}{\sigma_r(L)}. \tag{3.7}
\end{equation}
Normalization allows us to compare conditional shapes as a function of $L$ without confounding by scale.

\subsection{Binning and aggregation}

The aim is to estimate, for each lag $L$, the conditional expectation of the response given the push. We bin by the standardized push $z_p$ so shapes are comparable across lags. For each valid bin we report the response mean in raw units $r$ and in standardized units $z_r$.

We use one fixed grid for the standardized push $z_p$ that is shared by all $L$. Hasbrouck and Saar (2013) documented that low-latency strategies react to the current book state and that this reaction is state-dependent. Our use of a fixed grid in standardized pushes is a way to measure such state dependence uniformly across short and long lags. After winsorization and per-lag standardization, more than ninety-nine percent of $z_p$ mass lies within three standard deviations. We extend the grid to four standard deviations in both directions to capture informative tails while avoiding ultra-sparse extremes. The grid is
\begin{equation}
z_{p,\min} = -4.0,\qquad z_{p,\max} = 4.0,\qquad \Delta z_p = 0.025. \tag{3.8}
\end{equation}
This yields
\begin{equation}
N \;=\; \frac{z_{p,\max} - z_{p,\min}}{\Delta z_p} \;=\; 320 \quad \text{equal–width bins}. \tag{3.9}
\end{equation}
The step $\Delta z_p=0.025$ balances resolution and sampling error. With $2.6$ billion ticks and many lags, central bins contain thousands of observations per lag, which stabilizes bin means while preserving curvature in the push-response relation. Wider bins would smear structure, and narrower bins may create noisy averages in the wings.

For exact reproduction each standardized push $z_p$ maps to a bin index
\begin{equation}
j(z_p) \;=\; 1 + \left\lfloor \frac{z_p - z_{p,\min}}{\Delta z_p} \right\rfloor,\qquad j \in \{1,\dots,N\}, \tag{3.10}
\end{equation}
and each index has the bin center
\begin{equation}
z_{p,j} \;=\; z_{p,\min} + \Big(j - \tfrac{1}{2}\Big)\Delta z_p. \tag{3.11}
\end{equation}

We impose a minimum-support rule so that reported averages are not dominated by noise. A bin is valid at lag $L$ only if the number of admissible pairs in that bin reaches
\begin{equation}
n_{\min} \;=\; 200. \tag{3.12}
\end{equation}
With standardized responses this caps the standard error of a bin mean near $1/\sqrt{200}$ when $\sigma_r(L)=1$. Bins with fewer than $n_{\min}$ observations are omitted and never filled.

Let those $t$ for which $t-L$, $t$, and $t+L$ all lie in the same regular trading session be admissible anchors. For each lag $L$ and valid bin $j$ we report the count and the conditional response means
\begin{equation}
n(L,j) \;=\; \sum_{t} \mathbf{1}\!\left\{ j\!\big(z_p(t,L)\big)=j \right\}, \tag{3.13}
\end{equation}
\begin{equation}
\bar{z}_{p}(L,j) \;=\; \frac{1}{n(L,j)} \sum_{t:\, j(z_p)=j} z_{p}(L), \tag{3.14}
\end{equation}
\begin{equation}
\bar{z}_{r}(L,j) \;=\; \frac{1}{n(L,j)} \sum_{t:\, j(z_p)=j} z_{r}(L), \tag{3.15}
\end{equation}
The resulting object is a lag-by-bin grid with holes only where support is below $n_{\min}$, with no extrapolation and no imputation. 
Lillo, Mike, and Farmer (2005) showed that long memory in order signs can coexist with prices that are nearly efficient on average because the memory comes from order splitting. This is the reason to condition on both the lag \(L\) and the standardized push magnitude \(|z_p|\): if order splitting is present, the strength and even the sign of the conditional response can change with the horizon and with the size of the initiating move, so an unconditional autocorrelation would hide it. Therefore, the decomposition proposed is used to see the parts of the push-response curves that may contain long-memory effects.

\subsection{Symmetry, antisymmetry, and dominance}

Foucault, Kadan, and Kandel (2005) modeled the limit-order book as a market for liquidity in which the size of the liquidity-taking order and the current depth determine the price adjustment. This is the motivation for separating effects that depend on the size of the push from those that depend on its sign in the later decomposition.

Our object of interest is the conditional surface
\[
\overline{z}_r(L,z_p)\;= \frac{1}{n(L,j)} \sum_{t:\, j(z_p)=j} z_{r}(L)=\;\mathbb{E}\!\left[\,z_r \mid z_p,\,L\,\right], \tag{3.16}
\]
estimated on the bin grid described above. Near $z_p=0$ we can expect most pushes to concentrate tightly, and microstructure frictions might dilute information in the sign. If any predictability survives, it should emerge at moderate and large values of $|z_p|$. After visualizing the 3D push-response surface, we aim to separate magnitude-driven adjustment from sign-driven drift by decomposing the surface at each lag and absolute push.

For any lag $L$ and any pair of opposite bins $\pm j$ that both satisfy the support rule, define the even (symmetric) and odd (antisymmetric) parts:
\begin{equation}
S(L,|j|) \;=\; \tfrac{1}{2}\,\Big[\overline{z}_r(L,+j) + \overline{z}_r(L,-j)\Big], \tag{3.17}
\end{equation}
\begin{equation}
A(L,|j|) \;=\; \tfrac{1}{2}\,\Big[\overline{z}_r(L,+j) - \overline{z}_r(L,-j)\Big]. \tag{3.18}
\end{equation}
$S$ captures magnitude-conditioned adjustment that does not depend on the sign of the push. $A$ isolates sign-conditioned drift.

We introduce the simple measure with a unit-scale dominance index,
\begin{equation}
\rho(L,|j|)\;=\;\frac{A(L,|j|)}{|A(L,|j|)|+|S(L,|j|)|+\varepsilon},\qquad \varepsilon=10^{-12}. \tag{3.19}
\end{equation}
Values near $+1$ indicate local antisymmetry dominance. Values near $-1$ indicate local symmetry dominance. If either bin of the pair $\pm j$ fails the support rule, the cell is left blank.

However, to summarize by lag, we weight absolute-push pairs by usable information,
\begin{equation}
w(L,|j|)\;=\;\frac{n(L,+j)+n(L,-j)}{\sum_{|k|}\,[\,n(L,+k)+n(L,-k)\,]}, \tag{3.20}
\end{equation}
and define and report the lag-level dominance statistic
\begin{equation}
\rho(L)\;=\;\frac{\sum_{|j|}w(L,|j|)\,|A(L,|j|)|-\sum_{|j|}w(L,|j|)\,|S(L,|j|)|}{\sum_{|j|}w(L,|j|)\,|A(L,|j|)|+\sum_{|j|}w(L,|j|)\,|S(L,|j|)|}. \tag{3.21}
\end{equation}
This lies in $[-1,1]$ and is positive when antisymmetry dominates across $|z_p|$ at lag $L$.

Overall, response strength on the standardized scale is
\begin{equation}
M(L)\;=\;\sum_{|j|} w(L,|j|)\,\frac{|\overline{z}_r(L,+j)|+|\overline{z}_r(L,-j)|}{2}. \tag{3.22}
\end{equation}
The raw-unit analogue replaces $\overline{z}_r$ by the raw mean $\overline{r}(L,j)$ and, thus, yields $M_{\text{raw}}(L)$.

Uncertainty for $\rho(L)$ is attached by bootstrap resampling of symmetric bin pairs within each lag using probabilities $w(L,|j|)$. For $b=1,\dots,B$ we redraw pairs with replacement, recompute $\rho^{(b)}(L)$ from (3.21), and report pointwise 95\% bands from the empirical 2.5\% and 97.5\% quantiles. No unsupported cells are created by resampling.

\section{Results}\label{sec:results}

We visualize the conditional surface \(\overline{z}_r(L,z_p)\) defined by the bin means \(\overline{z}_p(L,j)\) on the common grid in \(z_p\) and across the two lag families. Each panel shows the same object from a different viewing angle so that both the across-lag evolution and the within-lag cross-sections are visible.

\begin{figure*}[!tbp]
  \centering
  \begin{minipage}[t]{0.48\linewidth}
    \centering
    \img[width=\linewidth,keepaspectratio]{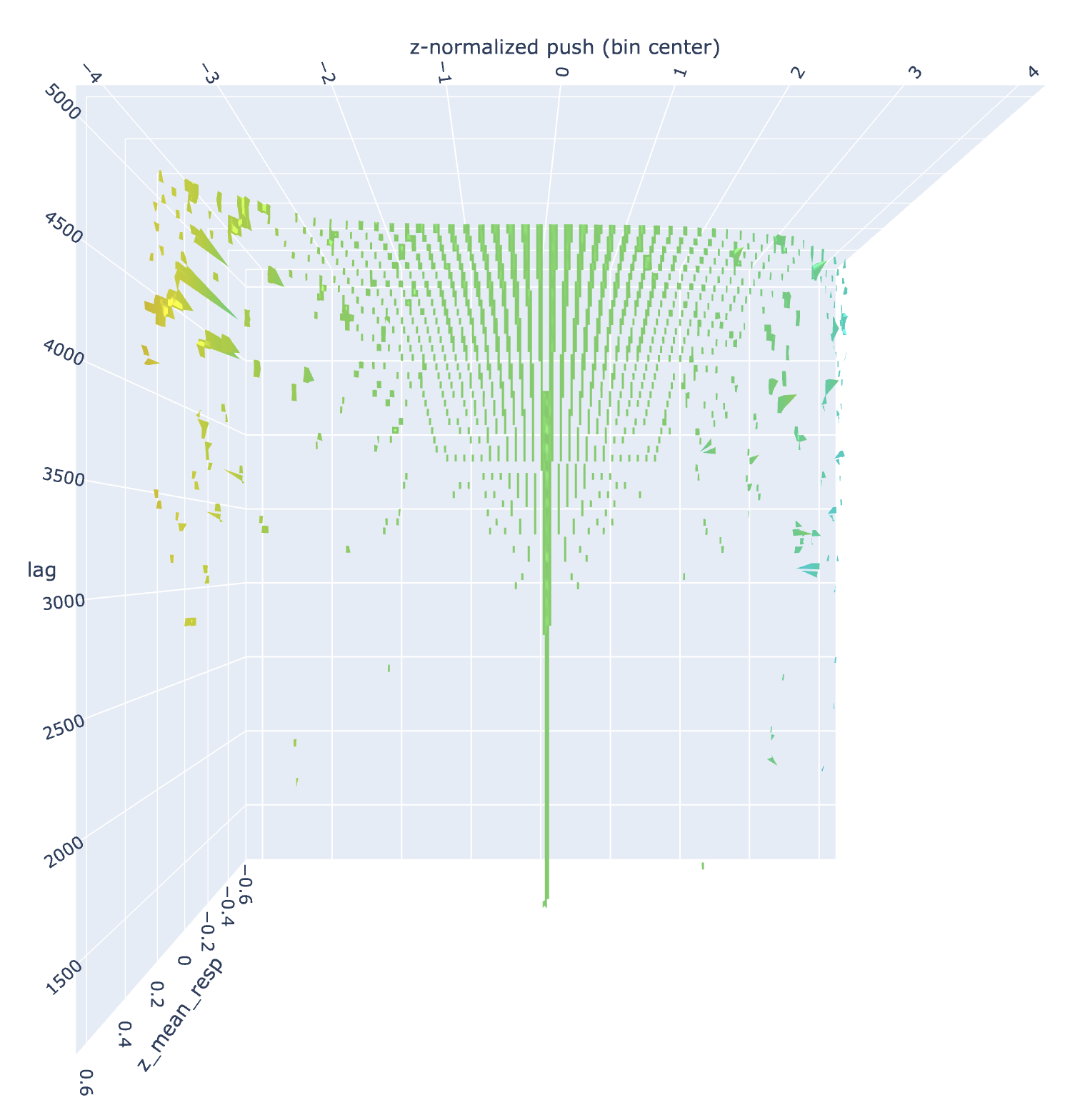}
    \subcaption{Short-lag surface, top view}
    \label{fig:short-top}
  \end{minipage}\hfill
  \begin{minipage}[t]{0.48\linewidth}
    \centering
    \img[width=\linewidth,keepaspectratio]{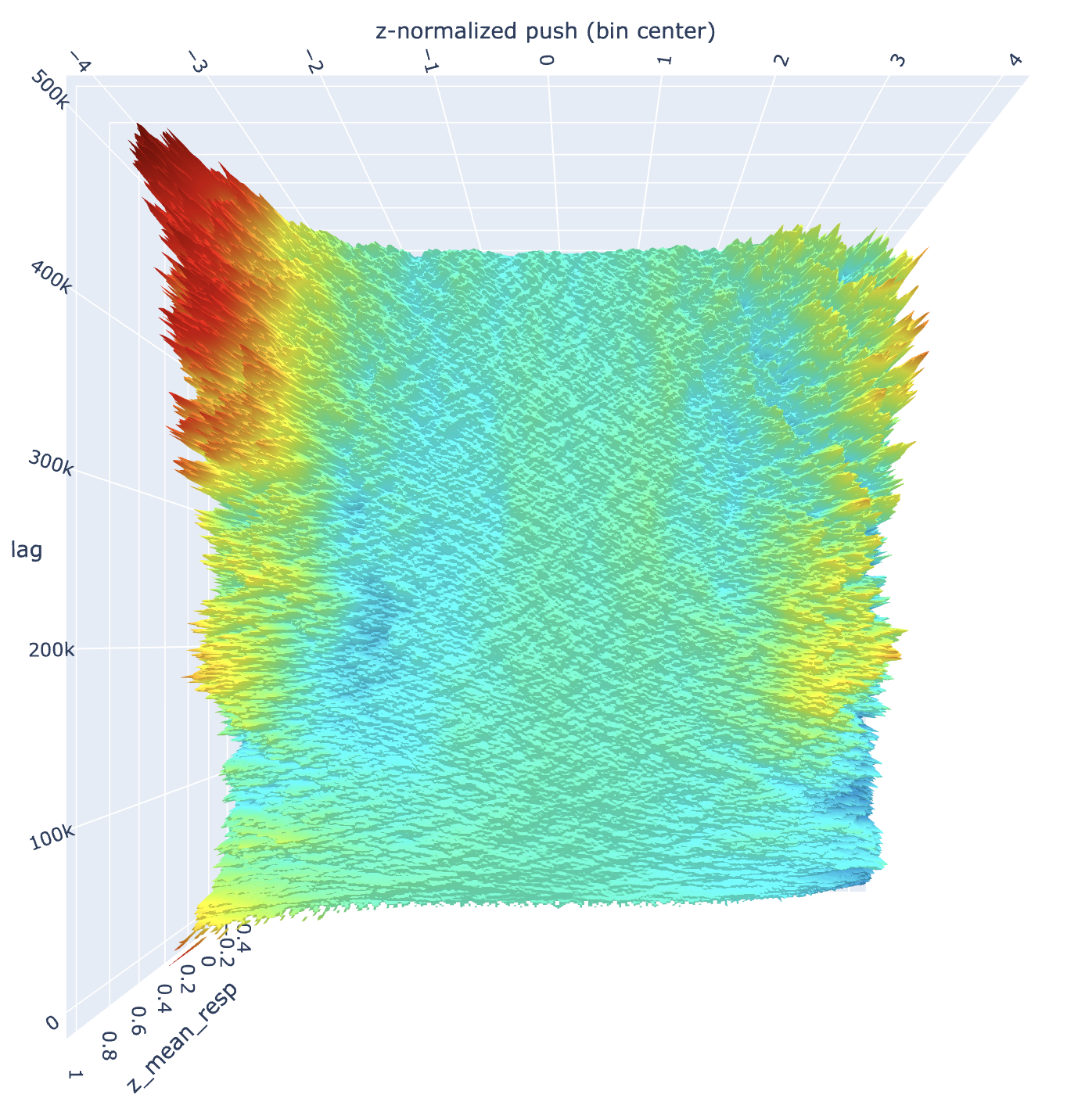}
    \subcaption{Long-lag surface, top view}
    \label{fig:long-top}
  \end{minipage}

  \vspace{0.7em}

  \begin{minipage}[t]{0.48\linewidth}
    \centering
    \img[width=\linewidth,keepaspectratio]{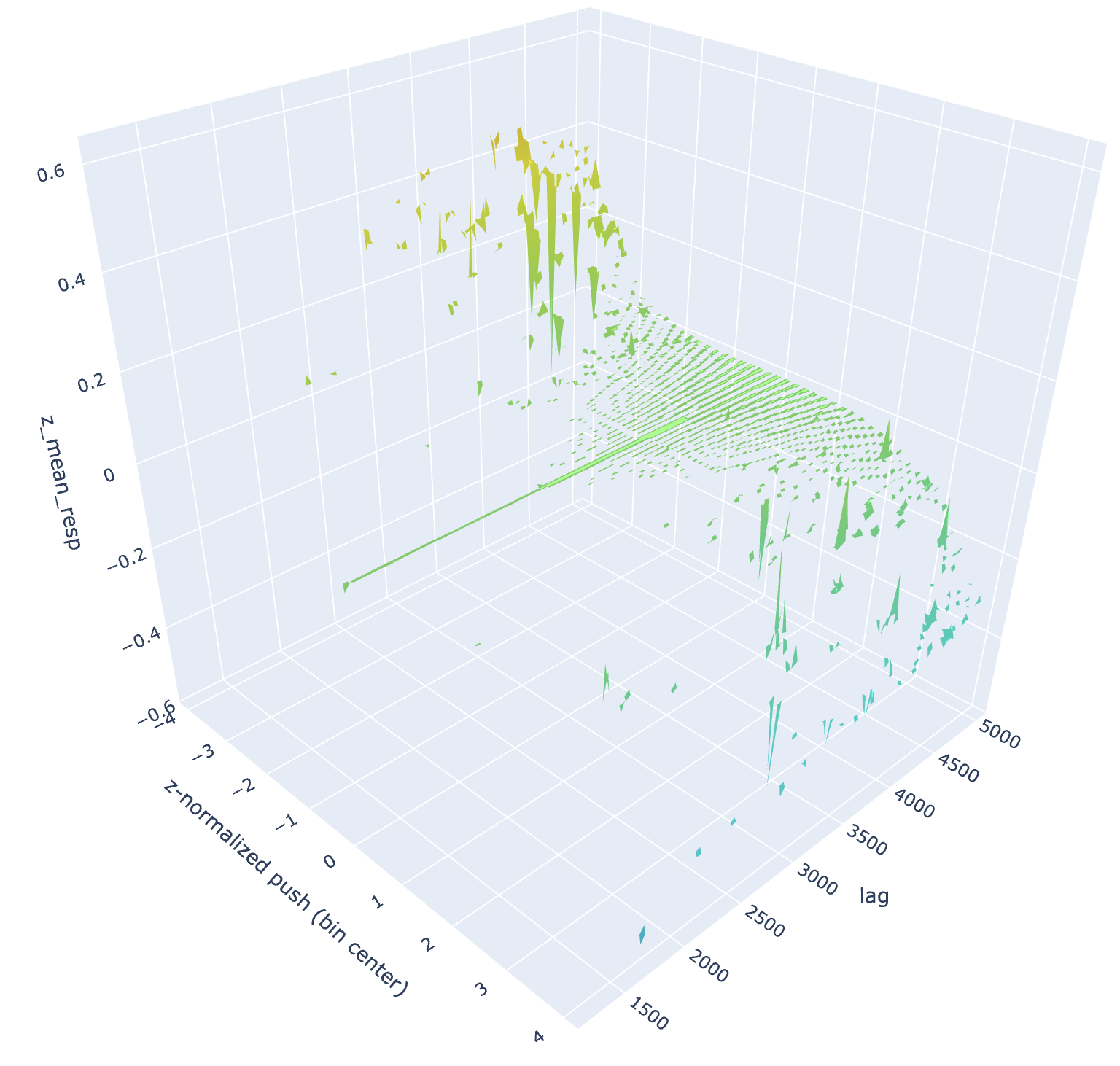}
    \subcaption{Short-lag surface, side view}
    \label{fig:short-side}
  \end{minipage}\hfill
  \begin{minipage}[t]{0.48\linewidth}
    \centering
    \img[width=\linewidth,keepaspectratio]{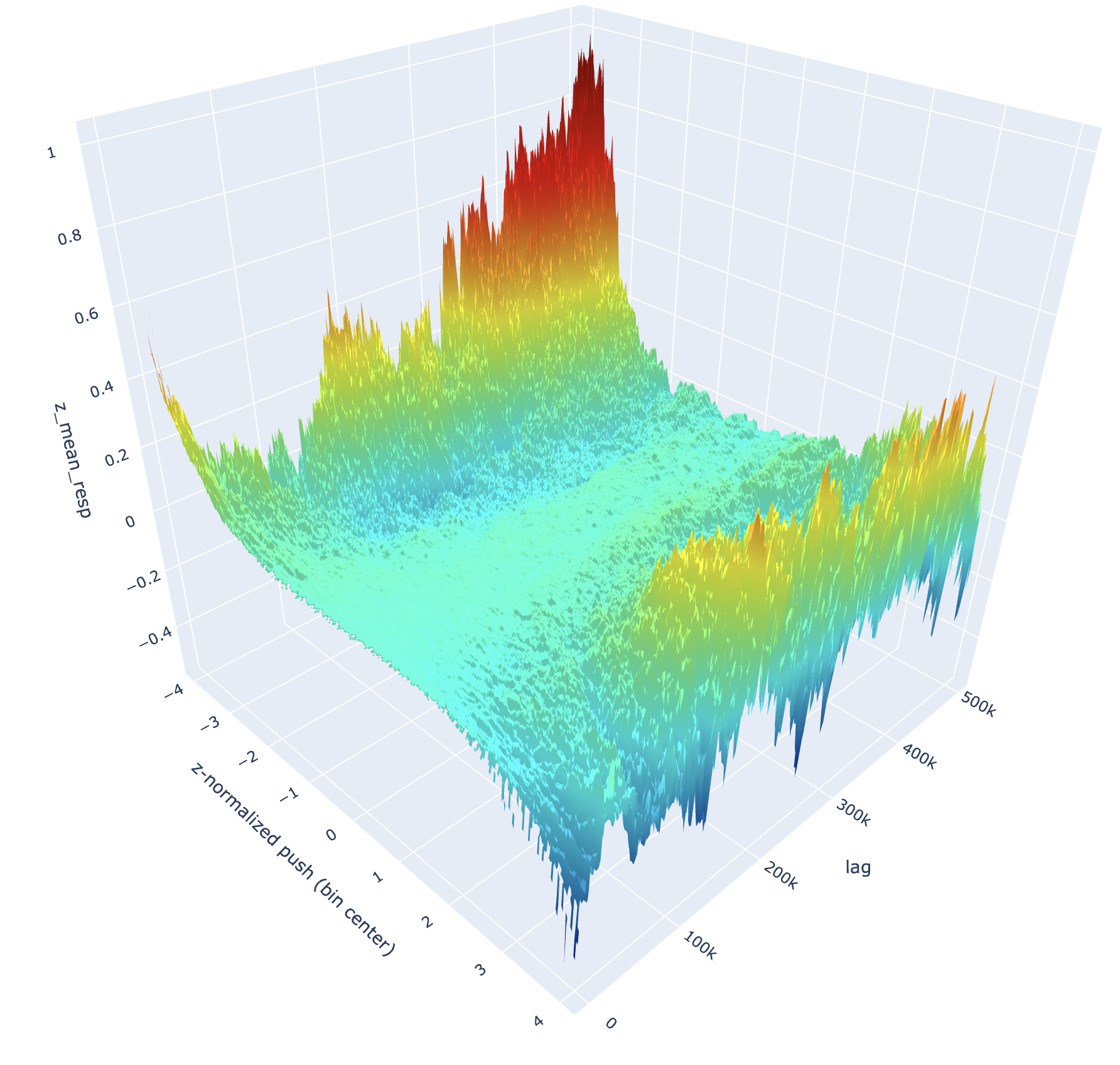}
    \subcaption{Long-lag surface, side view}
    \label{fig:long-side}
  \end{minipage}

  \caption{Push-response surfaces \(\overline{z}_r(L,z_p)\) for the short and long lag families. Top row: top views. Bottom row: side views.}
  \label{fig:surfaces}
\end{figure*}

The short-lag surface displays a narrow valley around \(z_p\approx 0\) where \(\overline{z}_p(L,j)\) is near zero and the support rule removes many tail bins. This region reflects the high density of small standardized pushes and the quick recycling of order flow at very short horizons. Where support appears in the wings, the response departs from zero and traces the onset of structure. The side view reveals that this structure turns on as soon as \(|z_p|\) moves past roughly one standard deviation. Within the short family the cross-sections in \(z_p\) remain thin and scattered. This is consistent with weaker tails at small \(L\) and with the fact that the same-session constraint trims many anchors for larger offsets even within the short grid.

The long-lag surface is dense and smooth: the top view shows a broad bowl around \(z_p\approx 0\) and a ridge that grows for \(|z_p|\in[1.3,3]\) as \(L\) increases from the lower end of the long grid toward the middle. The side view emphasizes that \(\overline{z}_r(L,j)\) changes sign across \(z_p=0\) in large regions, which signals an antisymmetric component. The amplitude of the surface rises with \(L\) up to a plateau for strong positive pushes: beyond that point the structure flattens, which indicates that predictability does not keep compounding at the far end of the grid once the information in the initiating push has diffused. However, for negative pushes, we observe an apparent lack of plateau on the negative side. This finding will be addressed later in the discussion. 

Also, on the lowest few thousand lags (roughly 5,000-150,000), the cross-sections of our push-response surface reproduce the classical “market mill” asymmetry reported by Leonidov, Trainin, Zaitsev, and Zaitsev (2006a, 2006b, 2006c, 2007, 2009): the reversal curves are evident, with positive pushes mapping to negative expected responses and negative pushes mapping to positive responses, strongest in the wings and attenuating near zero. This confirms their findings on our dataset, after which we generalize the effect across a dense lag grid and visualize it in 3D.

To separate magnitude-driven adjustment from sign-driven drift we use the symmetric and antisymmetric combinations from \((3.17)\)–\((3.18)\). The local dominance index \(\rho(L,|j|)\) in \((3.19)\) highlights which mechanism prevails at a given lag and magnitude.

\begin{figure*}[!tbp]
  \centering
  \img[width=0.89\textwidth,keepaspectratio]{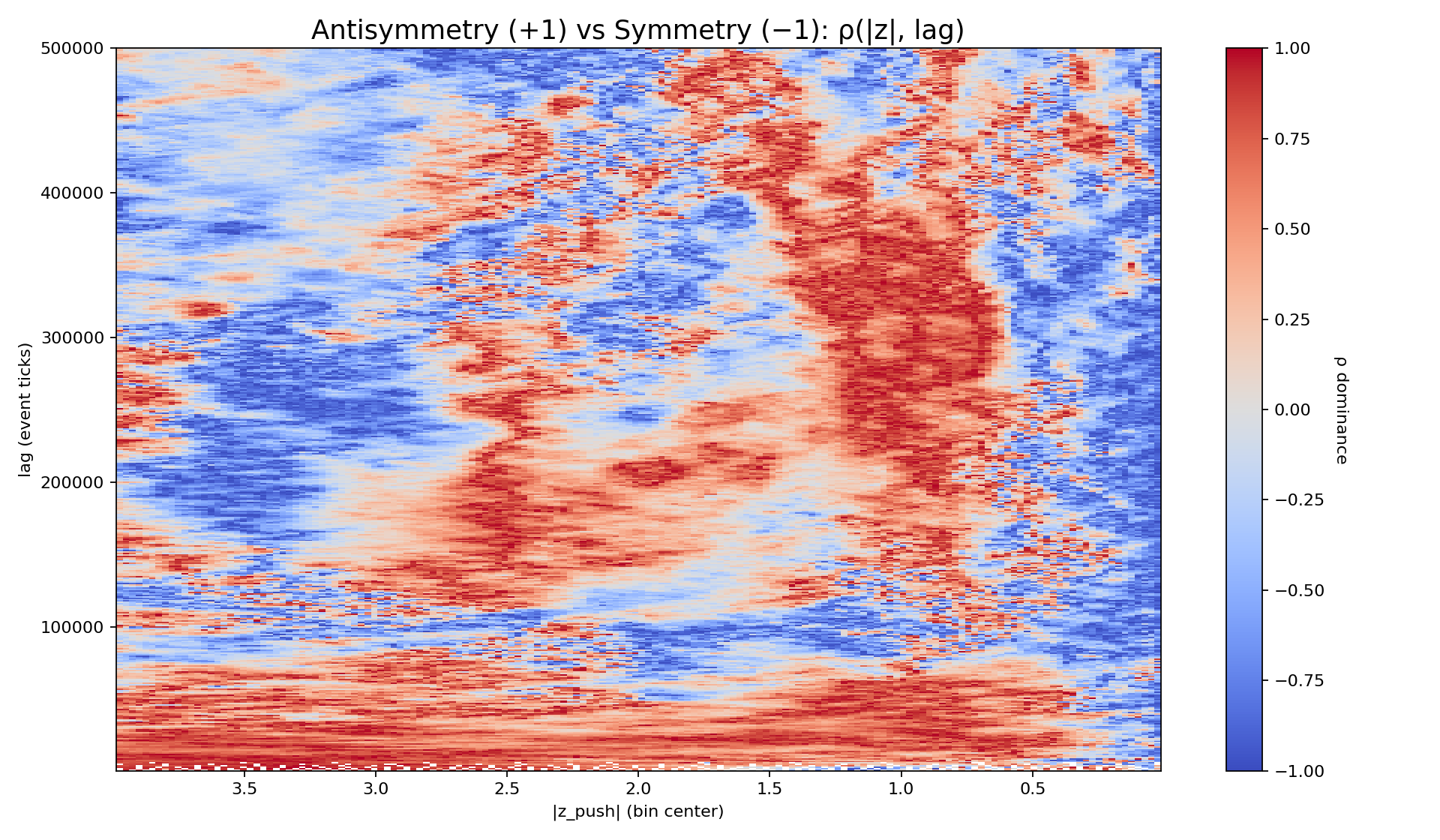}
  \caption{Dominance heatmap \(\rho(L,|z_p|)\) on the long-lag grid. Values near \(+1\) indicate antisymmetry dominance; values near \(-1\) indicate symmetry dominance. Cells are blank where at least one of the paired bins fails the support rule.}
  \label{fig:rho-heatmap}
\end{figure*}

Figure~\ref{fig:rho-heatmap} shows that near \(|z_p|\lesssim 0.7\) symmetry tends to dominate throughout, which matches the flat bowl in Figure~\ref{fig:surfaces}. For moderate magnitudes the heatmap turns positive over a wide band of lags. That band aligns with the ridge in the surface and implies that the sign of the initiating move carries through to the response at the same sign with similar magnitude. For larger lags, the dominance weakens and alternates in patches. This alternation is consistent with partial reversal of earlier continuation once inventory and liquidity effects unwind. The red-blue boundaries are not vertical: they drift in \(L\) as a function of \(|z_p|\). This drift means that the horizon at which sign information is strongest depends on the size of the initiating move, a signature of state dependence that would be obscured by an unconditional autocorrelation. Moreover, for the strongest pushes by magnitude, the heatmap is mostly negative, and additional evidence from the surface suggests that after strong initial price movements one can expect a positive response.

We summarize the overall strength of the surface through \(M(L)\) defined in \((3.22)\), and the aggregate sign-versus-magnitude balance through \(\rho(L)\) in \((3.21)\). Both statistics average across \(|z_p|\) with weights proportional to observed support.

\begin{figure*}[!tbp]
  \centering
  \img[width=0.95\textwidth,keepaspectratio]{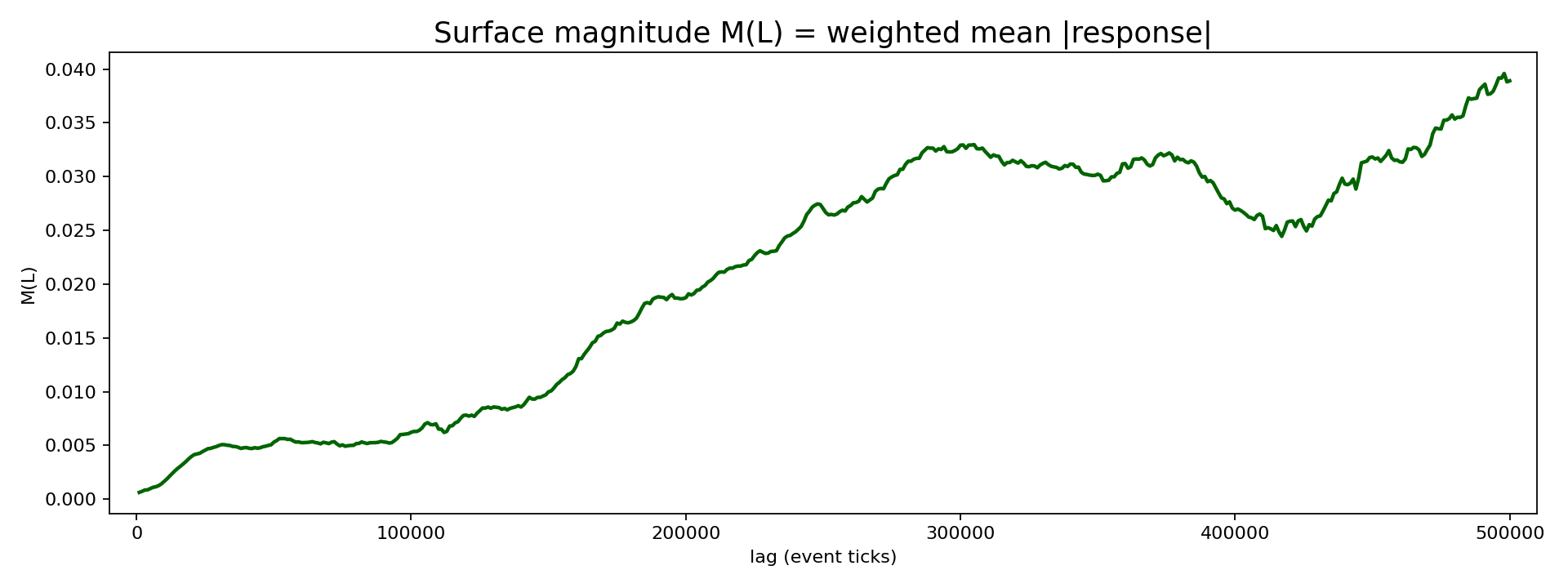}
  \caption{Surface magnitude \(M(L)\) on the standardized scale as a function of lag on the long grid}
  \label{fig:magnitude}
\end{figure*}

Figure~\ref{fig:magnitude} shows a monotone rise of \(M(L)\) from the bottom of the long grid into a broad plateau around the middle, followed by a mild increase. The initial rise is driven by the thickening of supported tail bins as \(L\) grows. The plateau indicates that the conditional response reaches a stable amplitude once the lag is long enough for order splitting and queue dynamics to propagate, but not so long that new shocks dominate the book. The late increase suggests that the standardized response picks up low-frequency components that are not tied to the most immediate microstructure effects.

\begin{figure*}[!tbp]
  \centering
  \img[width=0.95\textwidth,keepaspectratio]{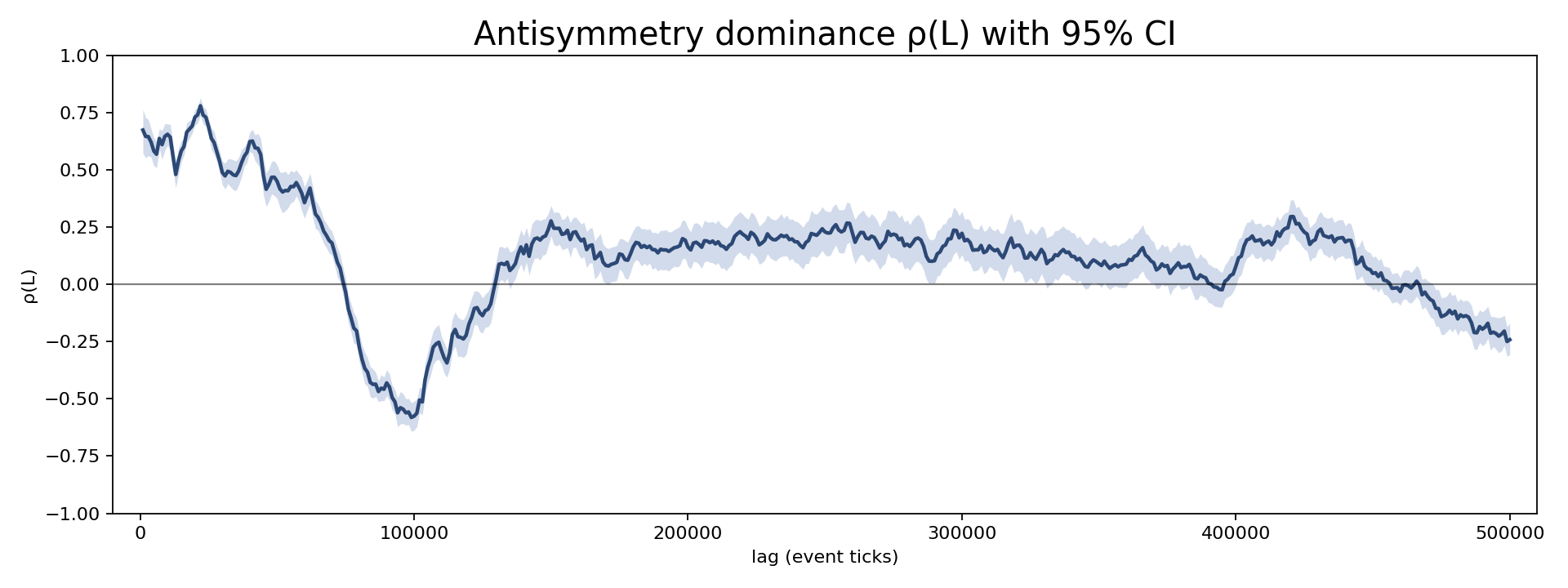}
  \caption{Aggregate dominance \(\rho(L)\) with pointwise 95\% bootstrap bands}
  \label{fig:rho-agg}
\end{figure*}

In Figure~\ref{fig:rho-agg}, $\rho(L)$ takes large positive values at the shortest lags, then turns negative around the transition from the short-lag grid to the early long-lag region, climbs back toward weakly positive levels across a broad middle range, and finally drifts slightly negative again at the longest lags. The positive region coincides with the red band in Figure~\ref{fig:rho-heatmap} and with the sign-changing cross-sections in the long-lag surface, which together indicate that the response tends to retain the sign of the push for moderate \(|z_p|\). The negative region is concentrated where the surface bowl is deepest and where \(S\) dominates; there the sign of the push matters less than its size. The confidence bands tighten where many symmetric pairs are supported and widen where wings thin out. The curve remains well inside its bands except at the very bottom of the grid where only a handful of pairs contribute and sampling variability is largest.

Taken together, the four panels in Figure~\ref{fig:surfaces} and the three summaries in Figures~\ref{fig:rho-heatmap}-\ref{fig:rho-agg} support the following reading. At very short horizons the conditional response is small and noisy after standardization; any apparent structure is local and sparsely supported. As lag increases the surface exhibits a clear ridge in \(|z_p|\) with a strong antisymmetric component, which implies sign-conditioned drift. The dominance of antisymmetry does not by itself fix momentum versus reversal, since that depends on the sign of \(\overline{z}_r(L,j)\) for positive \(z_p\). The long-lag surface shows both behaviors at different parts of the grid: momentum where \(\overline{z}_r(L,j)\) shares the sign of \(z_p\); reversal where the sign flips. The heatmap reveals that the balance between these regimes shifts with both lag and magnitude. For shorter lags, antisymmetry dominates. However, for longer lags, we see blue and red "rivers", where near-zero and highest pushes by absolute value have a strong symmetric component, but the ones that lie between 0.5 and 3 by absolute value represent higher antisymmetry. The magnitude curve confirms that these effects are not driven by a few isolated bins; they persist after averaging with weights that reflect observed support. The aggregate dominance curve shows that antisymmetry is materially present across a wide band of lags, then fades as the horizon stretches beyond the range where the initiating push still conditions the book.

\section{Further Discussion and Trading Implications}

This experiment shows systematic nonrandom structure in expected responses. The 3D surface and the antisymmetry heatmap together isolate where predictability concentrates across lags and magnitudes, rather than in unconditional averages.

First, larger pushes tend to trigger larger absolute responses. Whether the signal reads as trend following or contrarian depends strongly on the lag and push sign because the symmetric component is large for many horizons. This matches propagator intuition that the book adjusts with a state-dependent scale while the sign information waxes and wanes with horizon, consistent with Bouchaud, Gefen, Potters, and Wyart (2004) and with no-dynamic-arbitrage impact constraints in Gatheral (2010). Easley and O'Hara (1992) provided a dynamic model of how security prices adjust over time to imbalanced information arrivals. Also, Baltussen, Da, Lammers, and Martens (2021) document pervasive intraday momentum across global futures and link it to gamma-hedging demand from options market makers and leveraged ETFs. In our framework, such flow-driven continuation appears as same-sign push-response pockets with high antisymmetry at intermediate lags and moderate values of $|z_p|$. 

A practical implication is that long implementations often dominate short implementations in realized performance, both because shorting is costly and capacity-limited, and because many predictive regions of the surface -- where pushes have a higher magnitude, in particular -- exhibit symmetry that favors a positive response for both push signs. This is consistent with limits-to-arbitrage arguments in Shleifer and Vishny (1997) and with evidence that short-side anomalies are harder to monetize in practice in Stambaugh, Yu, and Yuan (2012). It also aligns with the superior long-only profitability of several well-documented factors, for example momentum in Jegadeesh and Titman (1993) when implemented with realistic short constraints, and with Novy-Marx’s (2013) result that a simple, scalable long-only exposure to gross profitability can deliver abnormal returns without relying on difficult-to-execute short legs.

Hendershott, Jones, and Menkveld (2011) showed on U.S. equities that algorithmic trading improves displayed liquidity but that the improvement is strongest when order flow is directional; in those states, price changes inherit the sign of the initiating flow. Menkveld (2013) described high-frequency market makers who continuously lean their quotes in the direction implied by recent trades. Brogaard, Hendershott, and Riordan (2014) quantified that such high-frequency traders contribute disproportionately to price discovery exactly at short horizons. Hendershott and Riordan (2013) extended this result to a broader sample of European stocks and confirmed that algorithmic liquidity supply is conditional on contemporaneous order flow. These four papers together imply that signed, short-horizon predictability is an equilibrium feature of modern order-driven markets, and our antisymmetry-dominant bands in $\rho(L,|z_p|)$ are an intraday, NBBO-level manifestation of the same mechanism.

Next, the surfaces reveal state dependence across magnitudes. Near small standardized pushes the conditional response is near zero and the heatmap tilts toward symmetry. As \(|z_p|\) grows into the range of one to three standard deviations and the ridge emerges, the antisymmetric component strengthens for wide bands of lags, and then attenuates at longer horizons. The ridge shifts with \(|z_p|\) rather than lining up at a single horizon, which indicates that the most informative horizon depends on shock size. This is consistent with long-memory order-flow and inventory dynamics in Lillo, Mike, and Farmer (2005) and with lagged order-imbalance effects in Chordia, Roll, and Subrahmanyam (2005). The plateau in the magnitude curve shows that conditional structure stabilizes rather than compounding without bound, which fits an efficient market that recycles microstructure pressure over time, echoing Poterba and Summers (1988) on eventual mean reversion at longer horizons.

Another feature of the push-response surface is the clear difference in magnitude of large initial moves: strong negative standardized pushes are followed, on average, by stronger positive conditional responses than the mirror positive pushes. Hasbrouck (1991) explained that when sell-side pressure leaves dealers with long inventory, quotes are revised upward to restore inventory targets, which mechanically produces a positive follow-up move. Biais, Hillion, and Spatt (1995) also documented that the limit-order book replenishes asymmetrically after one side is hit, with buy orders arriving more aggressively after heavy selling. Our lag-resolved estimates are consistent with this microstructure mechanism: shocks that deplete the bid elicit a faster and larger restoring adjustment than shocks that deplete the ask, so the conditional response surface is steeper on the negative side.

With regard to the application of this finding, one can build a systematic strategy that maps realized pushes at a fixed lag or a small set of lags to the corresponding conditional responses from the standardized surface and then converts these responses into an expected P\&L distribution. The standardized surface lets traders compare horizons on a common scale and neutralize the timing effect that would otherwise push the selection toward very long lags purely because raw variance grows with lag. 

This framework can drive two distinct uses: a signal that selects statistically supported regions of the surface after transaction costs, and a risk metric that attributes realized P\&L to lag-specific conditional responses.

For experiment design we used tick data to compare short- and long-lag behavior on a unified backbone. For trading, a coarser bar construction may be preferable to align with execution horizons and to keep holding periods within operational limits when lags number in the hundreds of thousands in event time. Extending the analysis with explicit transaction-cost overlays should also improve trading implementability.

Finally, SPY is highly liquid and efficient, so the same methodology applied across instruments should reveal richer structure where microstructure frictions are more pronounced. The right unit of analysis is the asset-specific surface: identify lags where the post-cost mean-standardized response is materially different from zero within its confidence interval, then restrict trading to those pockets.

\section{Concluding Remarks}

In this paper, we examined whether consecutive price increments in a highly liquid equity instrument, SPY, exhibit conditional dependencies that persist after lag-specific standardization. Rather than testing for autocorrelation in aggregate, we constructed a lag-resolved push-response decomposition that preserves directional and magnitude information across a wide range of horizons. The hypothesis that such dependencies exist -- and are not purely random noise -- was strongly supported by empirical structure across lags and push magnitudes.

We found that the expected forward return conditioned on a standardized past return is not zero for many lag-magnitude combinations. The conditional surface revealed nontrivial geometry: at very short lags, the response is noisy and concentrated near zero, consistent with strong informational efficiency. However, as lag increases, a structured ridge emerges in the surface that traces a zone of predictability. This ridge is not symmetric about zero but shifts with lag and push size, indicating that the horizon at which information propagates depends on the size of the initial shock.

By decomposing the surface into its symmetric and antisymmetric components, we were able to isolate magnitude-conditioned and sign-conditioned effects. The symmetric component dominated for small pushes and extreme lags, while the antisymmetric component -- responsible for directional drift -- peaked at intermediate lags and moderate pushes. This structure was confirmed quantitatively by the lag-aggregated counterpart, which traced a region of antisymmetry dominance across the central lag band. These results are consistent with partial momentum and partial reversal, depending on the location in the lag-magnitude space.

A further contribution of our analysis is the documented asymmetry in conditional adjustments following large negative price moves. Across lags, sharp downside pushes are followed by stronger positive conditional responses than upside pushes of comparable magnitude. This is consistent with liquidity-restoration dynamics in limit-order markets, where aggressive selling tends to deplete the bid and leave intermediaries with inventory imbalances, prompting subsequent replenishment and upward price adjustment that is more pronounced than the analogous response after large buys. The asymmetry underscores that predictability is not merely directional but conditional on the market state produced by the preceding shock, further supporting the push-response decomposition approach.

This methodologically distinct decomposition offered clarity where unconditional correlation measures typically fail. The traditional view is that SPY is near efficient. Our analysis shows that localized predictability exists in narrow support-stable pockets. These findings align with microstructure theories involving inventory risk, asymmetric liquidity provision, and delayed order flow adjustment discussed in the stated literature.

From a practical perspective, the surface we construct serves as both a predictive signal and a risk attribution tool. Because the conditioning is on standardized push values, expected responses can be interpreted in normalized units and directly compared across lags. This enables construction of lag-aware, variance-adjusted trade filters, which can selectively activate strategies when the observed push falls within empirically predictive zones. It also provides a framework for explaining realized P\&L in terms of forecasted conditional responses.

Future extensions should move beyond SPY to instruments with wider spreads, thinner depth, or more fragmented liquidity. We also see scope for incorporating realized transaction costs, order book state variables, and time-of-day controls. Bogousslavsky (2021) decomposes returns into intraday and overnight components and shows that cross-sectional predictability loads differentially on these windows. As our estimates are confined to regular trading hours, his results motivate a day-night segmentation of our surface to test whether the ridge persists, attenuates, or flips across the close. Further, one could apply machine learning to learn nonlinear mappings of pushes to responses. For instance, Lucchese, Pakkanen, and Veraart (2024) use deep-learning models on order-book states to show that short-horizon predictability is ubiquitous and horizon-dependent. Thus, their multi-horizon design can be a machine-learning generalization of our nonparametric mapping from standardized pushes to responses and a way to probe cross-asset universality of the lag-resolved structure. Ultimately, these findings offer the view that the path to equilibrium is structured, lagged, and observable, while being full of local anomalies that should be further explored.

\nocite{*}
\vspace{0.5em}
\bibliographystyle{agsm}
\bibliography{references}

\end{document}